\UseRawInputEncoding
\documentclass[conference]{IEEEtran}
\IEEEoverridecommandlockouts
\usepackage{cite}
\usepackage{amsmath,amssymb,amsfonts}
\usepackage{graphicx}
\usepackage{textcomp}
\usepackage{xcolor}
\usepackage{booktabs} % For better horizontal lines
\usepackage{calc}     % For calculations in LaTeX
\usepackage{siunitx}
\usepackage{amsmath}
\usepackage{algorithm}
\usepackage{algpseudocode}
\usepackage{algorithmicx}
\usepackage{algpseudocode}
\usepackage{url}

\def\BibTeX{{\rm B\kern-.05em{\sc i\kern-.025em b}\kern-.08em
    T\kern-.1667em\lower.7ex\hbox{E}\kern-.125emX}}
\begin{document}

\title{A Novel Topology Recovery Method for Low Voltage Distribution Networks\\
}

\author{\IEEEauthorblockN{Sina Mohammadi, Van-Hai Bui, Wencong Su}
% \thanks{S. Mohammadi, V.-H. Bui, and W. Su are with the Department of Electrical and Computer Engineering, University of Michigan– Dearborn, Dearborn, MI 48128, USA.}% <-this % stops a space
\thanks{This paper is accepted for presentation in 2025 IEEE PES General Meeting.}% <-this % stops a space
\IEEEauthorblockA{\textit{Department of Electrical \& Computer Engineering} \\
\textit{University of Michigan-Dearborn}\\
Dearborn, USA \\
sinamo, vhbui, wencong@umich.edu
}
\vspace{-5mm}
% \and
% \and
% \and
% \and
% \and
}

\maketitle

\begin{abstract}

Low voltage distribution networks (LVDNs) suffer from limited visibility due to sparse or nonexistent measurement systems, leaving distribution network service providers with incomplete data. Maintenance activities, such as transformer upgrades and power line replacements, sometimes go undocumented, leading to unmonitored topology changes. This lack of oversight hinders network optimization, fault detection, and outage management, as utilities cannot fully monitor or control the system. With the rise of electric vehicles, having an accurate understanding of LVDN topology is crucial to avoid infrastructure damage from potential overloads. This paper introduces a method to reconstruct LVDN topology using incremental voltage and current changes from smart meters at customer endpoints. The approach identifies and maps network topologies with high accuracy, overcoming limitations of prior methods by discarding unrealistic assumptions. Specifically, it addresses grids with fewer than three pole connections and employs an AC power flow model over simplified DC approximations. Simulations across diverse different configurations validate the method’s effectiveness in accurately reconstructing LVDN topologies, enhancing real-world applicability.

\end{abstract}


\section{Introduction}
Distribution networks deliver electricity to residential and commercial loads. The topology of distribution network may alter due to multiple reasons including switching scenarios or outages. Knowing the grid structure is very important to detect faults, outage management or power flow optimization. Nowadays, many metering systems are deployed to capture the real time power system behavior. In transmission level, there are lots of measurement systems that can monitor the grid accurately. However, in distribution grids, the lack of enough measurement devices, manual switches and undocumented changes cause to have an unobservable grid. In LVDNs, finding the true topology is more challenging since installed metering systems are sparse or not available. Thus, there is a need to figure out the current true topology of the grid to do optimal operations. A distribution transformer along with multiple poles and conductors are utilized to energize the residential loads which is hidden to utilities in case of topology changes. Therefore, knowing the pole places and conductor connections are crucial to make grid-aware optimal decisions. Nowadays, smart meters are deployed in many houses to capture real time data for electricity consumption. These smart meters are able to acquire voltage, current, active/reactive power in short time intervals such as every 15 minutes \cite{9569740}. This provides enough data to observe load profiles, billing history, voltage and topology estimation as well.

There are a plenty of past works that investigates the topology estimation in distribution networks \cite{deka2023learning,dalavi2024review}. The voltage-based topology correction is proposed in \cite{luan2015smart} where the initial topology extracted from GIS maps are corrected. The highly correlated voltage profiles show the shorter distances between load points. The work \cite{koivisto2012clustering} presented a method to monitor the status of switches in distribution network. Thus, by knowing the status changes of switches, the topology of the system can be obtained. However, all the switching scenarios are not feasible in real grids causing high computational burden utilizing this method. Besides, the initial topology is needed as a prior knowledge. In \cite{10251548}, a graph theory-aided method is provided to decrease the exhaustive search for feasible switching scenarios. Then a quadratic programming is utilized to find the system topology. However, the DC power flow is considered in the formulations which is not realistic. Modeling the grid topology as a graph and encodes it via Prufer decoding is investigated in \cite{soumalas2017data}. The voltage sensitivity matrix and power deviations are considered to indicate the path resistance between any arbitrary pair of nodes. Due to small values of sensitivity matrix, this method is prone to noise and transients. Another voltage sensitivity-based method is proposed in \cite{10025832} where the voltage of the distribution transformer could be varied and the paths between any pair of load nodes and transformer is estimated to reconstruct the grid topology. Mixed integer linear programming is used in \cite{tian2015mixed} for topology estimation. The nodal voltages and line flows should be available. However, in real distribution systems nodal voltage metering is sparse and line flow measurement are almost absent. 

There is a group of works that utilized statistical indexes-based power flow calculations to estimate the topology for distribution networks \cite{deka2015structure1,deka2015structure2,deka2016learning,deka2017structure,park2020learning}. Full availability of nodal measurements along with correlated power injections are the some of unrealistic assumptions in \cite{deka2015structure1,deka2015structure2}. In \cite{deka2017structure}, partial observability of the system is considered. In other words, some of the nodes in the graph-represented structure is hidden and unmonitored. Recursive grouping algorithm (RGA) is used to determine the network topology step by step. The RGA groups correlated observed nodes step by step by defining a parent and reconstruct the topology till no grouping is possible. In the presented works, DC power flow is utilized to formulate the problem which is not accurate in real world scenarios. In addition, the RGA only applicable in such grids that all hidden nodes have the degree equals or greater than three. The node degree represents the number of power lines that are connected to a specific node. This constraint is not realistic in radial distribution networks where nodes with degree two is present.

In \cite{6687939}, the Ybus estimation considering DC power flow is proposed for power system topology estimation. The Ybus matrix encodes all the topology information and connectivity among all grid nodes. While the paper goal is not distribution network, it provides an efficient mathematical formulation to investigate topology estimation with only power injection data via Ybus estimation. In \cite{8456535}, the Lasso algorithm is implemented to estimate the Ybus of distribution grid. The Ybus encodes the nodal connections as well as line impedances, so, if one can estimate the Ybus, the topology also can be estimated. Since the Ybus matrix is a sparse matrix, Lasso is more efficient than least square method which enables the presence of zero entries in Ybus estimation process. However, the full nodal voltage and current measurements should be available to apply Lasso for topology estimation. In \cite{10620006}, an enhanced Ybus estimation is proposed that DC, DC coupled and AC power flow equations are considered to estimate for Ybus and is more realistic, while the all nodal measurements are needed. 

In this paper, a simple approach leveraging joint incremental voltage and current signals is proposed for the topology recovery of LVDNs. In contrast with numerous prior studies, this method relies solely on data from smart meters and accounts for a partially observable LVDN, enhancing its applicability to real-world scenarios where full network observability is often unachievable. Unlike simplified DC power flow models, this approach employs AC power flow, offering a more accurate representation of the characteristics in LVDNs. The proposed methodology accommodates scenarios with fewer than three pole connections and relaxes the assumption of uncorrelated nodal power injections, allowing it to operate under conditions where nodal dependencies exist. With a sufficient number of measured samples, this method demonstrates the capability to achieve a 100\% accuracy rate in topology recovery, underscoring its robustness and effectiveness in practical distribution network applications.

The rest of the paper is organized as follows: In section II the graph represented LVDN is introduced and the proposed method along with a simple example is discussed considering the assumptions in the proposed method. Section III includes the simulation results for the proposed method with four different test system. Finally, section IV concludes the paper.

\begin{figure}[t]
    \centering
    \includegraphics[width=0.8\linewidth]{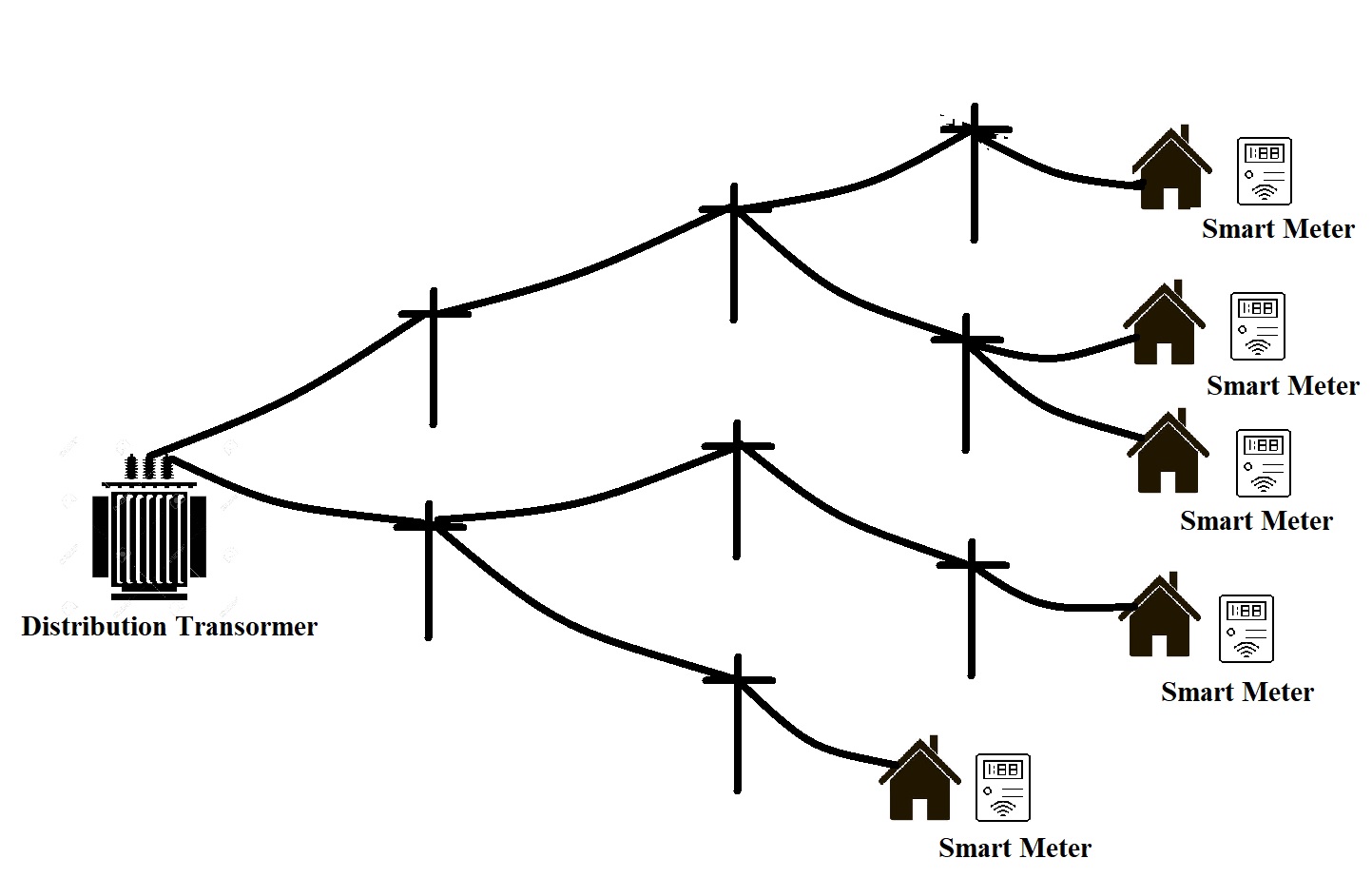}
    \caption{A typical LVDN with 100\% smart meter coverage of residential loads}
    \label{LVDN}
    \vspace{-3mm}
\end{figure}

\section{Distribution Network Estimation}

\subsection{Distribution Network As a Graph}
The LVDN can be modeled as a graph \( G = (V, E) \), where \( V \) represents the set of nodes (poles) and \( E \) represents the set of edges (power lines). Since the LVDN operates in the radial mode, \( G \) is a tree \( T \) without loops. In this tree \( T \), the leaf nodes \( L \) correspond to the customers and are the only observable nodes, forming a set of observed nodes \( O \subseteq V \). The unobserved or hidden nodes \( H \) are defined as \( H = V \setminus O \), resulting in a hidden tree structure. The supporting poles, which are not connected to loads and lack measurement devices, are included in \( H \). The objective is to estimate this hidden tree using only available smart meter data from the observed nodes. Presented approach involves determining the distance between the nodes using a distance-based grouping method to infer the network topology. Fig. \ref{LVDN} shows a LVDN with low voltage transformer, supporting poles and customers that are all equipped with smart meters \cite{10400804}.

\begin{figure}[b]
\centering
\begin{tabular}{c}
    \includegraphics[width=0.9\linewidth]{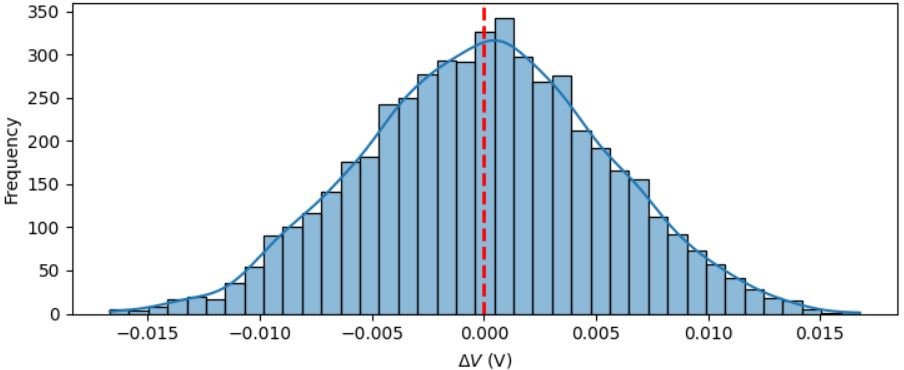}\\
    (a)\\
    \includegraphics[width=0.9\linewidth]{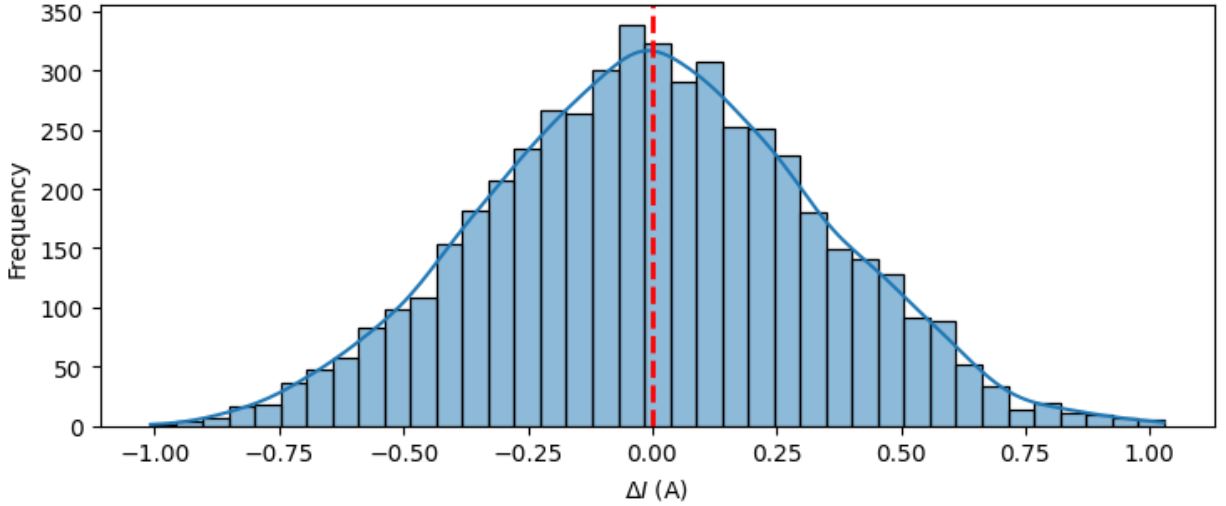}\\
    (b)\\

\end{tabular}
   \caption{For the 11-node test system, histograms for Node 2 display the distribution of voltage incremental changes ($\Delta V$) and current changes ($\Delta I$) across the system.}
   \label{incremental profiles}
   \vspace{-5mm}
\end{figure}

\begin{figure}[t]
    \centering
    \includegraphics[width=0.4\linewidth]{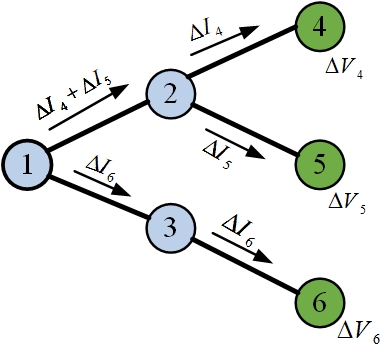}
    \caption{Schematic of a 6-node grid used in the proposed method derivation. Green nodes represent observable nodes, while light blue nodes indicate hidden nodes.}
    \label{simple 6 node}
    \vspace{-3mm}
\end{figure}

\subsection{Proposed Method for Topology Estimation}
In this paper topology of LVDN is estimated by utilizing the incremental change of voltage and current signals of smart meters. As mentioned in \cite{8456535}, we assume that all internal/hidden nodes in the grid neither have no power injection or consumption. In other words, hidden poles are just supporting poles to carry electricity from low voltage distribution transformer to the loads. In distribution networks we have $R>>X$ for power lines \cite{7736082} which implies that the power lines can be modeled as purely resistive lines. Besides, the physical distances between any two nodes in LVDNs are almost equal. For instance, the physical distance for rural areas is 30-40 meters. As a result, the electric distances, which are directly associated with physical distances, could be obtained via the conductor electric characteristics and line length. Thus, in this paper, all the power lines are assumed to be same with known resistance. Now, the problem of topology estimation is converted to finding node connectivity and is not a joint estimation problem to find both connections and line impedances.

The voltage and current data profile in power systems have irregular distribution \cite{8456535} in LVDNs. As shown in \cite{7741545}, increment change of voltage magnitude data ($\Delta V$), whose realization at time t is $\Delta V[t] = V[t] - V[t - 1]$ and current magnitude ($\Delta I[t] = I[t] - I[t - 1]$) have Gaussian distribution. Based on Fig. \ref{incremental profiles}, the simulation results validated that incremental change of voltage and current have normal distribution. We used these properties in our proposed method to apply statistical tools for finding the topology estimation. 

Consider Fig. \ref{simple 6 node} as it depicted a simple topology with only 6 nodes and 5 segments. In this structure, nodes 4, 5 and 6 are leaf nodes and 1, 2 and 3 are hidden nodes in which node 1 is the root node. Firstly, the magnitude of $\Delta V$s and $\Delta I$s data of leaf nodes are acquired. All collected data are then scaled using standard scaling for better data representation. Standard scaling Normalizes the dataset by adjusting values to have a mean of zero and a standard deviation of one. Scaling can mitigate the negative impact of small values in the collected data without changing trends and dependencies.

Utilizing $\Delta V$s we have a set of multivariate Gaussian distribution data. The correlation matrix can be derived as:
\begin{equation}
    \Sigma = \frac{1}{n-1} (X - \mu)^T (X - \mu)
    \label{eq.1}
\end{equation}

Where \(X\) represents the data matrix with observations and variables, \(\mu\) is the mean vector of the data, and \(n\) denotes the number of observations. Based on the correlation matrix, one can obtain the precision matrix that shows the dependencies between any pair of variables (here nodes) and is as follows:
\begin{equation}
   \Theta = \Sigma^{-1} 
\end{equation}

Accordingly, the distance matrix that enumerates the closeness or farness of the variables can be written:
\begin{equation}
     D_{ij} = \frac{1}{|\Theta_{ij}|}
    \label{eq.2}
\end{equation}  

where \(\Theta_{ij}\) is the \((i, j)\)-th element of the precision matrix, and \(|\Theta_{ij}|\) ensures non-negative distances, reflecting the strength of conditional dependencies. 

\begin{figure}[t]
\centering
\begin{tabular}{ccc}
    \includegraphics[width=0.35\linewidth]{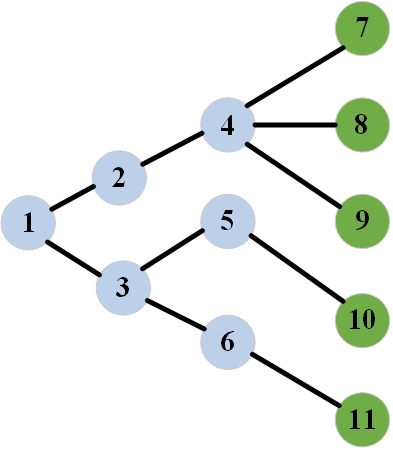} &  \includegraphics[width=0.45\linewidth]{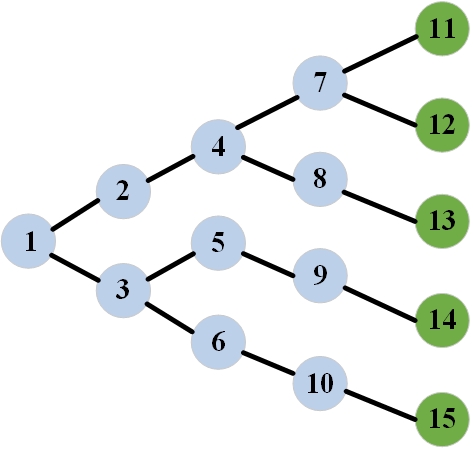}\\
    (a) & (b)\\
    \includegraphics[width=0.4\linewidth]{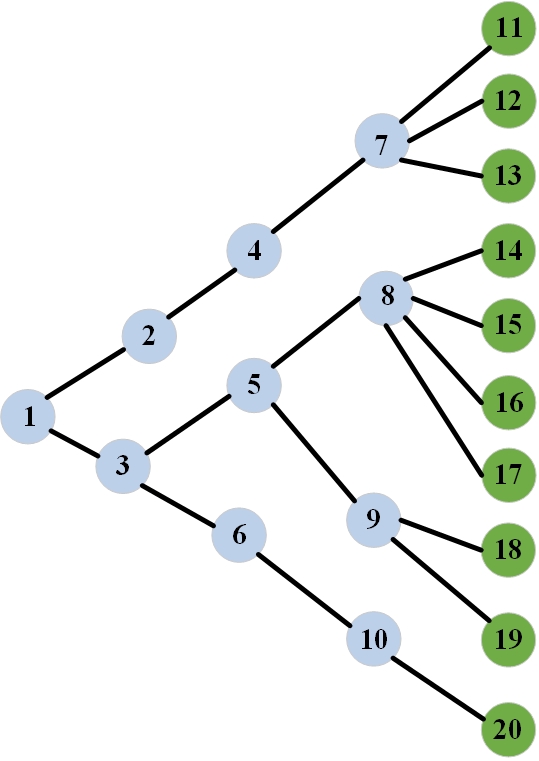} & \includegraphics[width=0.5\linewidth]{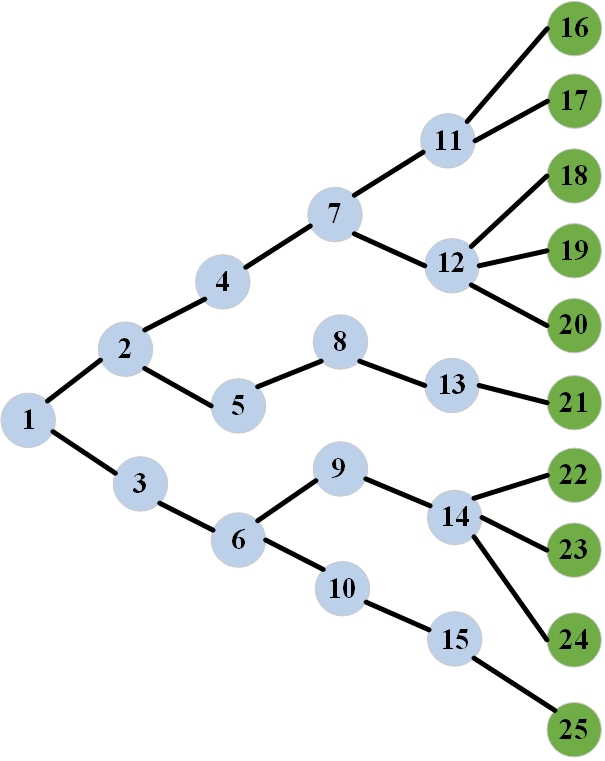}\\
    (c) & (d)\\

\end{tabular}
   \caption{Graph representations of test systems for simulation results. Light blue nodes indicate internal hidden poles, while green nodes represent observable consumer loads with smart meters. (a) 11-node system, (b) 15-node system, (c) 20-node system, (d) 25-node system.}
   \label{test systems}
   \vspace{-5mm}
\end{figure}

Now, by knowing the distances, based on the incremental voltage data of leaf nodes, the first grouping step can be done to find the first layer of hidden nodes as the parents of leaf nodes. As it is possible to have more than two leaf nodes as siblings, it is necessary to consider a threshold for grouping aim. In this paper the threshold distance value is selected as 0.2 and derived from multiple simulations of four test systems which are presented later in simulation results. For various topologies, the 11-node test system exhibits higher sensitivity toward grouping buses 7, 8, and 9 in the initial step, representing the worst-case scenario in our study. Consequently, by setting the threshold at 0.2, grouping more than two nodes is consistently ensured across all test configurations. The leaf nodes with distance lower than 0.2 are all grouped with one common parent. Otherwise, each ungrouped leaf node is connected to single child parent node. 

As the first step done, based on Fig. \ref{simple 6 node} we have:

\begin{equation}
\label{dv1}
    \left\{
    \begin{aligned}
        \Delta V_2 &= \Delta V_4 + R\Delta I_4 \\
        \Delta V_2 &= \Delta V_5 + R\Delta I_5 
    \end{aligned}
    \right.
\end{equation}

\begin{equation}
\label{dv2}
    \Delta V_3 = \Delta V_6 + R\Delta I_6
\end{equation}

Based on Eq.4, there are two values for $\Delta V_2$ and one value for $\Delta V_3$. One can easily use the mean value to obtain $\Delta V_2$. In this step, again the grouping is done based on the obtained $\Delta V$ values for hidden nodes which now are observable. As shown in Fig. \ref{simple 6 node}, the root node as the parent of node 2 and 3 is reached and there is no possible further grouping in this step. Thus, this is the termination pint of the proposed method where all nodes are grouped. This procedure can easily extend to any arbitrary network with observable leaf nodes and hidden interval nodes to reconstruct the topology only with leaf nodes data. The summarized steps for topology recovery are outlined in Algorithm \ref{alg:topology}.

\begin{algorithm}[t]
\caption{Proposed Topology Recovery Algorithm}\label{alg:topology}

\begin{algorithmic}[1]

\small
\Statex \textbf{Input:} $\Delta V$ and $\Delta I$ of leaf nodes (smart meters)
\Statex \textbf{Output:} Recovered topology of the LVDN

\State Data collection of smart meters (voltage and current data)
\State Extract the incremental voltage and current data and scale the data using standard scaling.
\State Group the the closest leaf nodes based on $\Delta V$ distances considering the threshold value.
\State Finding the $\Delta V$ values for first layer of hidden nodes based on the leaf node data diffusion and group the closest nodes.
\State Repeat step 3 and 4 to capture all the hidden nodes and ultimately root node where no further grouping is possible.
\State Recover the full topology of the LVDN.
\end{algorithmic}
\end{algorithm}

\section{Simulation Results}
To validate the proposed method, four different test systems are considered. Since there are no widely recognized benchmark topologies for low voltage grids, we consider different generated structures with various number of leaf/hidden nodes. In this way, we try to consider different possible nodal layers and connections in generated grid topologies. As the LVDNs have lower number of nodes compared to primary networks due to line power losses, the grids with 11, 15, 20 and 25 nodes with 5, 5, 10 and 10 observable leaf nodes are considered respectively (see Fig. \ref{test systems}). The line-to-line voltage of all systems is considered as 400 volts and node to node length is considered 30 meters, which is realistic value in urban distribution networks. The conductor resistance per ohm is considered to be 0.03$\si{\ohm}/km$ and the overall line (segment) resistance is equal to $9\times10^{-3} \si{\ohm}$. The load profiles, are extracted and interpolated from real residential data which are available for 24 hours \cite{hourlyLoadProfiles}. In fact, we solved multiple power flows based on the leaf node load changes to collect the incremental variables which are required for topology estimation. The backward-forward sweep algorithm is utilized to generate the power flow solutions which is a type of AC power flow calculations. Once the data is collected, one can apply the proposed methodology to recover topology step by step from leaf nodes to root node.

\begin{figure}[t]
    \centering
    \includegraphics[width=0.95\linewidth]{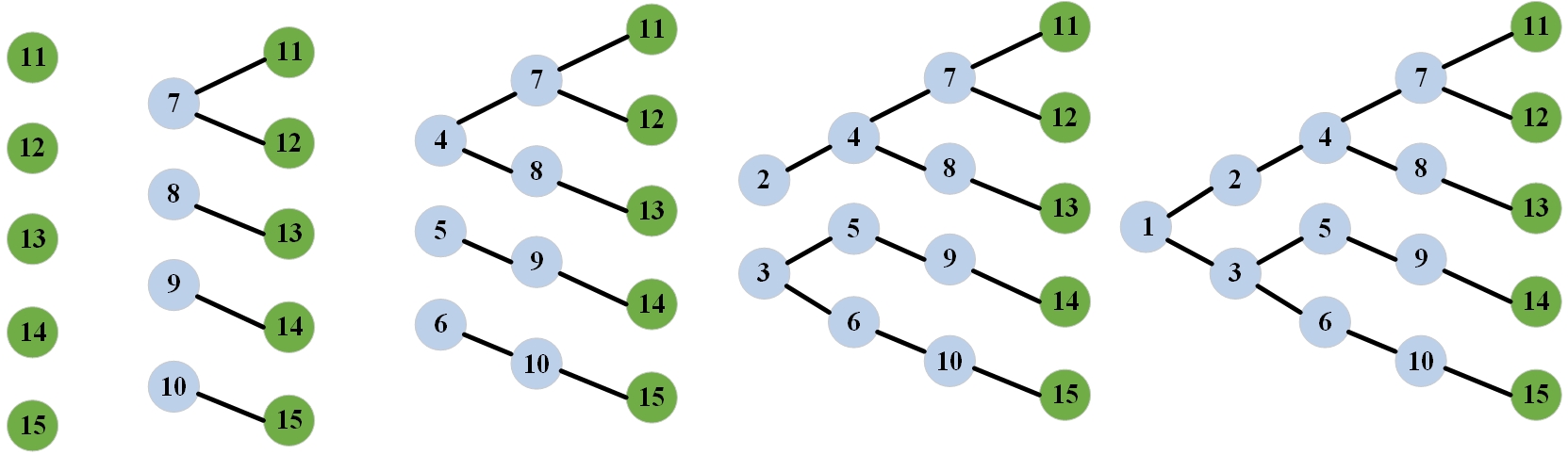}
    \caption{Step-by-step grid reconstruction process for a 15-node test system. The sequence illustrates the progressive stages of reconstructing the grid topology.}
    \label{step by step}
    \vspace{-3mm}
\end{figure}

\begin{table}[t]
    \centering
    \caption{Step-by-step Reconstruction of Different Test Systems Considering The $\Delta V$ Distances}
    \begin{tabular}{p{2.85cm}p{\dimexpr\columnwidth-3cm-2\tabcolsep-1em\relax}}
        \toprule
        \textbf{Test System} & \textbf{Details} \\ \midrule
        11 node (1000 samples) & 
        \textbf{Step 1:} Nodes 7, 8, 9 (Distance=0.1766) \\
        & \textbf{Step 2:} Nodes 10, 11 (Distance=1.1511) \\
        & \textbf{Step 3:} Nodes 7, 8, 9, 10, 11 (Distance=34.8143) \\ \midrule

        15 node (1000 samples) & 
        \textbf{Step 1:} Nodes 11, 12 (Distance=0.0676) \\
        & \textbf{Step 2:} Nodes 11, 12, 13 (Distance=0.5986) \\
        & \textbf{Step 3:} Nodes 14, 15 (Distance=1.8121) \\
        & \textbf{Step 4:} Nodes 11, 12, 13, 14, 15 (Distance=24.3685) \\ \midrule

        20 node (15000 samples) & 
        \textbf{Step 1:} Nodes 18, 19 (Distance=0.0715), Nodes 11, 12, 13 (Distance=0.0960), Nodes 14, 15, 16, 17 (Distance=0.1064) \\
        & \textbf{Step 2:} Nodes 14, 15, 16, 17, 18, 19 (Distance=1.9473) \\
        & \textbf{Step 3:} Nodes 14, 15, 16, 17, 18, 19, 20 (Distance=6.6794) \\
        & \textbf{Step 4:} Nodes 11, 12, 13, 14, 15, 16, 17, 18, 19, 20 (Distance=51.2522) \\ \midrule

        25 node (5000 samples) & 
        \textbf{Step 1:} Nodes 16, 17 (Distance=0.0338), Nodes 18, 19, 20 (Distance=0.0368), Nodes 22, 23, 24 (Distance=0.0528) \\
        & \textbf{Step 2:} Nodes 16, 17, 18, 19, 20 (Distance=1.3764) \\
        & \textbf{Step 3:} Nodes 22, 23, 24, 25 (Distance=2.8200) \\
        & \textbf{Step 4:} Nodes 16, 17, 18, 19, 20, 21 (Distance=13.6254) \\
        & \textbf{Step 5:} Nodes 16, 17, 18, 19, 20, 21, 22, 23, 24, 25 (Distance=31.2434) \\ \bottomrule
    \end{tabular}
    \label{tab:reconstruction_steps}
\end{table}

Fig. \ref{step by step} shows the step-by-step grid reconstruction for 15 node test system with 1000 samples of incremental voltage and current data of leaf nodes. As evident, buses 11 and 12 are grouped in the first step, while other nodes are just connected to a single child parent node. In this stage, the first hidden layer and nodes are recovered. Then, the leaf nodes data are distributed throughout the hidden nodes by considering the circuit analysis and KCL and KVL rules (see Eq. \ref{dv1} and \ref{dv2}). Thus, the obtained incremental voltage data are used to group the closest nodes iteratively till the root node is recovered and no further grouping is possible. 

Table \ref{tab:reconstruction_steps} represents the number of steps, the grouped nodes in each step and the associated distances. As shown, the number of required steps for topology recovery is equal to number of hidden layers for all four test systems. Besides, the selected threshold guarantees the correct grouping in the first recovery step.
\begin{figure}[t]
\centering
\begin{tabular}{cc}
    \includegraphics[width=0.45\linewidth]{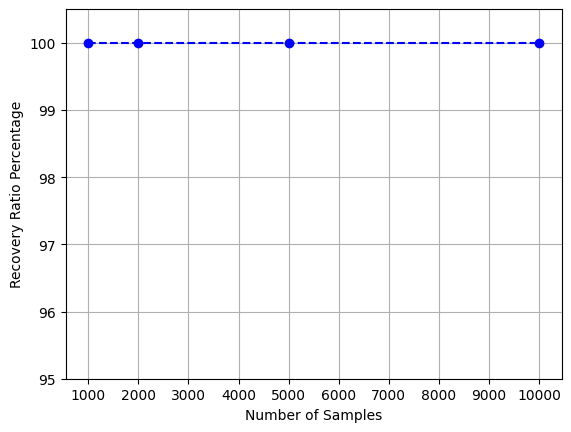} &  \includegraphics[width=0.45\linewidth]{recovery1.png}\\
    (a) & (b)\\
    \includegraphics[width=0.45\linewidth]{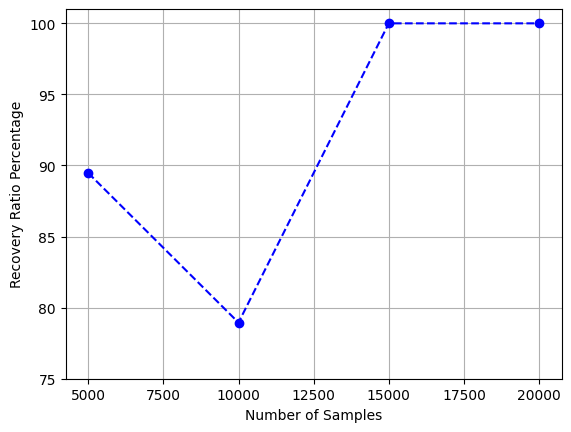} & \includegraphics[width=0.45\linewidth]{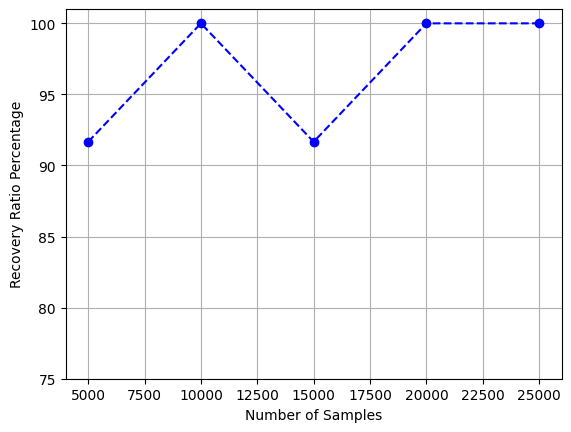}\\
    (c) & (d)\\

\end{tabular}
   \caption{Recovery ratio percentages for test systems of varying sizes. Subplots show results for (a) 11-node, (b) 15-node, (c) 20-node, and (d) 25-node systems, illustrating the algorithm's performance across different grid complexities.}
   \label{recovery}
\end{figure}
The provided results in Table I showcase the topology reconstruction with the samples required for 100\% accurate topology estimation. 

It is useful to examine the method's recovery error against the number of samples provided. Figure \ref{recovery} illustrates the recovery ratio for the four test systems analyzed. The recovery ratio represents the proportion of correctly estimated lines to the total lines in the graph, serving as a measure of estimation accuracy. For systems with 11 and 15 nodes, the proposed method successfully recovers the true topology with only 1,000 samples. However, the 20- and 25-node systems require more data to achieve 100\% accuracy in topology estimation. Nevertheless, zero error in topology estimation is achievable for both the 20- and 25-node grids.

\vspace{-1mm}
\section{Conclusions}
% This paper introduces a novel method for topology recovery in low-voltage distribution networks, utilizing only smart meter data from residential customers. The approach leverages residential load characteristics to identify hidden nodes and connections through the information entropy of observed leaf nodes. Circuit analysis propagates this information across the grid, inferring incremental voltage data for hidden nodes, which is then used to group nodes by electrical distance, reconstructing the grid topology. Simulations confirm the method's accuracy, achieving 100\% topology recovery with minimal measurement data. Future work includes adapting this approach for unknown line impedances and meshed grid structures.

This paper presented a novel method for topology recovery in low-voltage distribution networks using only smart meter data. The proposed approach leverages prior knowledge of residential load characteristics to determine hidden nodes and their nodal connections through the information entropy of observed leaf nodes. By incorporating circuit analysis, this information is propagated throughout the grid to infer incremental voltage data for hidden nodes. This data is then used to group nodes based on their distances, thereby reconstructing the overall grid topology. Simulation results validate the effectiveness of the proposed method, demonstrating that it can accurately recover the network topology with 100\% precision. This method significantly enhances the efficiency of topology estimation, particularly for practical low-voltage networks. Future research directions include extending the proposed approach to handle scenarios involving unknown line impedances, as well as adapting it to meshed grids.

\bibliographystyle{ieeetr}
\bibliography{ref}

\end{document}